\documentclass[9pt,conference]{IEEEtran}

\usepackage{cite}
\usepackage{amsmath,amssymb,amsfonts}
\usepackage{algorithmic}
\usepackage{graphicx}
\usepackage{textcomp}
\usepackage{xcolor}
\def\BibTeX{{\rm B\kern-.05em{\sc i\kern-.025em b}\kern-.08em
    T\kern-.1667em\lower.7ex\hbox{E}\kern-.125emX}}

\usepackage[tableposition=top]{caption}
\usepackage{subcaption}

\usepackage{listings}

\usepackage{pifont}
\usepackage[hidelinks]{hyperref}

\definecolor{codegreen}{rgb}{0,0.6,0}
\definecolor{codegray}{rgb}{0.5,0.5,0.5}
\definecolor{codepurple}{rgb}{0.58,0,0.82}
\definecolor{backcolour}{rgb}{0.95,0.95,0.92}

\usepackage{booktabs}
\usepackage{multirow}

\usepackage{enumitem}
\setlist[itemize]{leftmargin=*}
\setlist[enumerate]{leftmargin=*}

\usepackage{scrextend}

\begin{document}

\title{Joint Semantic Knowledge Distillation and Masked Acoustic Modeling for Full-band Speech Restoration With Improved Intelligibility}

\author{\IEEEauthorblockN{Xiaoyu Liu}
\IEEEauthorblockA{\textit{Dolby Laboratories} \\
}
\and
\IEEEauthorblockN{Xu Li}
\IEEEauthorblockA{\textit{Dolby Laboratories} \\
}
\and
\IEEEauthorblockN{Joan Serr\`a}
\IEEEauthorblockA{\textit{Dolby Laboratories} \\
}
\and
\IEEEauthorblockN{Santiago Pascual}
\IEEEauthorblockA{\textit{Dolby Laboratories} \\
}
}

\maketitle

\begin{abstract}
Speech restoration aims at restoring full-band speech with high quality and intelligibility, considering a diverse set of distortions. MaskSR is a recently proposed generative model for this task. As other models of its kind, MaskSR attains high quality but, as we show, intelligibility can be substantially improved. We do so by boosting the speech encoder component of MaskSR with predictions of semantic representations of the target speech, using a pre-trained self-supervised teacher model. Then, a masked language model is conditioned on the learned semantic features to predict acoustic tokens that encode low level spectral details of the target speech. We show that, with the same MaskSR model capacity and inference time, the proposed model, MaskSR2, significantly reduces the word error rate, a typical metric for intelligibility. MaskSR2 also achieves competitive word error rate among other models, while providing superior quality. An ablation study shows the effectiveness of various semantic representations.
\end{abstract}


\section{Introduction}
\label{sec:introduction}

Speech restoration (SR) aims at restoring full-band speech with high quality and intelligibility from a corrupted signal~\cite{liu2022voicefixer, serra2022universal, chen2023gesper, koizumi2023miipher}. Compared with conventional speech enhancement (SE) that typically employs discriminative modeling based on regression to remove noise~\cite{zhao2022frcrn} and, at most, reverb~\cite{defossez2020real, li2021simultaneous}, SR addresses a diverse set of tasks including those that are generative in nature, such as bandwidth extension, packet loss concealment, etc. For both SR and SE, a common finding is that the improved perceptual quality after processing may not translate to improved intelligibility, typically measured by the word error rate (WER) of automatic speech recognition (ASR) systems, since removing distortions could alter the phonetic content~\cite{wang2023speechx, koizumi2023miipher, scheibler2024universal, e3net}. Actually, the processed speech may even have a higher WER than that of the corrupted speech~\cite{wang2023speechx, koizumi2023miipher, e3net}. One solution is to condition the model on the text transcription of the corrupted speech, which, however, may not be available during both training and inference~\cite{wang2023speechx, koizumi2023miipher, le2024voicebox}. Another approach optimizes the model with an additional ASR-related loss~\cite{scheibler2024universal, e3net}, but pre-training the ASR model requires large transcribed datasets. 

\begin{figure*}[t]
  \centering
  \includegraphics[width=0.95\textwidth]{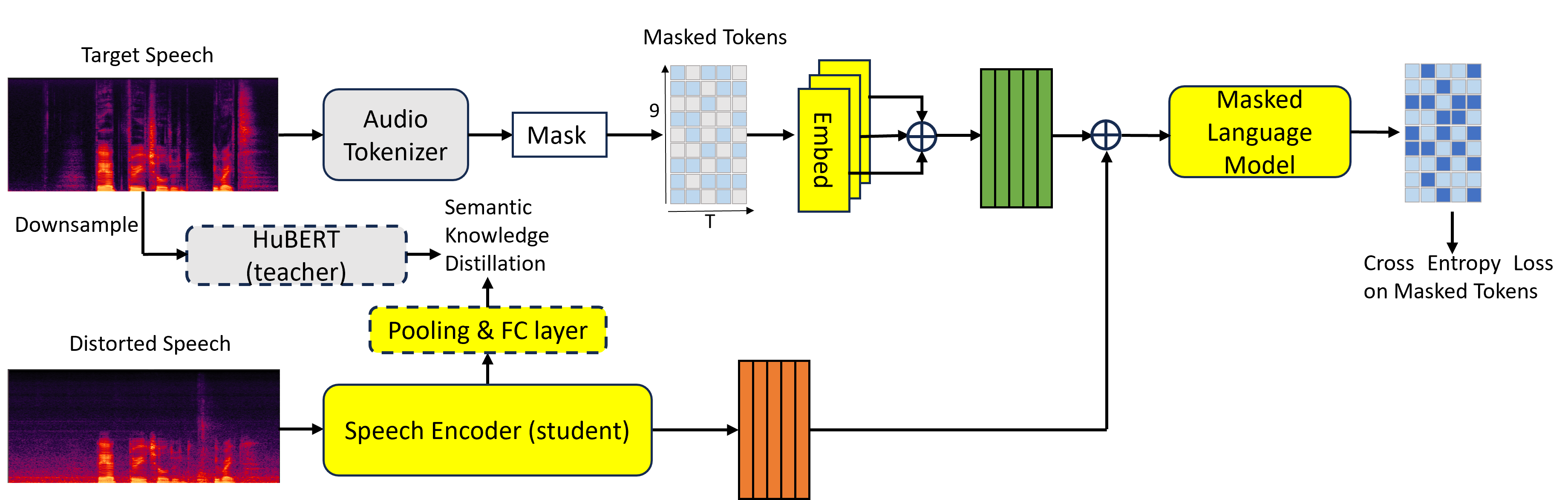}
  \caption{The MaskSR2 architecture. The yellow blocks are trainable and the grey ones are frozen. Compared with MaskSR, MaskSR2 introduces the semantic knowledge distillation during training as a loss function for the speech encoder. During inference, the dashed blocks are discarded, reducing MaskSR2 to MaskSR, thus maintaining model size and inference time.}
  \vspace{-2ex}
  \label{fig:masksr2}
\end{figure*}

In this paper, we propose MaskSR2 (Fig.~\ref{fig:masksr2}), which significantly reduces the WER without relying on transcribed data. MaskSR2 is based on our previous work MaskSR, a full-band (44.1\,kHz) SR system that holistically performs denoising, dereverberation, declipping, and bandwidth extension under a masked token modeling paradigm~\cite{li2024masksr}. MaskSR2 improves upon MaskSR by introducing semantic knowledge distillation (KD) in the speech encoder component. During training, given the STFT of a corrupted speech signal, the speech encoder predicts semantic representations of the target speech, extracted using a pre-trained HuBERT model~\cite{hsu2021hubert}. This teacher model encodes semantic (phonetic) patterns learned through self-supervised learning (SSL), which removes the need for transcribed audio. The KD is imposed by a loss function upon the speech encoder, jointly optimized with the rest of the system. Meanwhile, the generative model is conditioned on the learned semantic features from the hidden layer of the encoder to predict randomly-masked acoustic tokens of the target speech. During inference, the HuBERT teacher is discarded, thus reducing the compute cost of MaskSR2 to the one of MaskSR (same model capacity and inference time). Iterative sampling is performed on the output distribution to generate the target acoustic tokens, which are then converted to a waveform by a pre-trained audio (de)tokenizer.

We get inspiration from previous text-guided speech/audio synthesis research, which shows the importance of semantic modeling~\cite{agostinelli2023musiclm, kharitonov2023speak, liu2024audioldm, dong2023clipsonic}. These systems usually consist of two stages, text-to-semantic and semantic-to-acoustic synthesis, each requiring separate training and iterative inference. For SR, since the corrupted speech provides a stronger condition than a text prompt, we are able to jointly train the speech encoder and the generative model, and then run iterative sampling only on the latter. Another SR work~\cite{serra2022universal} also fuses a speech encoder with a generative model, but the encoder is optimized only on spectral targets, such as the STFT and features derived from it. We show that the semantic KD is an effective choice to improve intelligibility. SELM~\cite{wang2024selm} trains a language model to translate discrete noisy SSL tokens to clean ones, thus performing denoising with lower WER. But this framework requires storing and running a (typically large) SSL model during inference, which is not needed by MaskSR2. Also, the generated speech quality is sensitive to the codebook size of the discrete SSL tokens~\cite{wang2024selm}. We avoid this issue by exploring continuous HuBERT features as the teacher for the speech encoder. Overall, MaskSR2 reduces the WER between 19\,\% and 38\,\% relative to MaskSR, and achieves a competitive WER when compared to strong regression models trained to optimize waveform or spectrum alignment (thus tending to produce lower WER than generative models). MaskSR2 also outperforms various models in terms of quality. Samples are available on our demo page\footnote{\url{https://masksr.github.io/MaskSR2/}}.

\section{MaskSR2}
\label{sec:masksr2}

\subsection{Speech Encoder}
\label{sec:encoder}

MaskSR2 improves the speech intelligibility on top of MaskSR by introducing semantic KD to the speech encoder. Given a speech signal $x$ sampled at 44.1\,kHz, distorted by the studied artifacts (including noise, reverb, low bandwidth, and clipping), the speech encoder first computes the power-law compressed magnitude STFT, $X^{0.3}$, using a window and hop size of 2048 and 512 samples, respectively. Next, a 1025-channel 1-D batch normalization layer normalizes the magnitude of each frequency bin. Then, a fully connected (FC) layer projects the normalized features to $d$ dimensions, and a stack of self-attention transformer blocks learn a sequence of latent embeddings. The output from the last transformer block is used to condition the generative model. MaskSR does not employ any auxiliary loss to train the speech encoder, whereas MaskSR2 aims at learning semantic content without transcribed data. For this purpose, we consider learning from a HuBERT-base model\footnote{\label{note1}\url{https://github.com/facebookresearch/fairseq/tree/main/examples/hubert}}~\cite{hsu2021hubert} pre-trained on the 960-hour LibriSpeech dataset~\cite{panayotov2015librispeech}. The self-supervised pre-training encodes semantic information by predicting masked pseudo-phonetic labels derived only from audio. The resulting intermediate features show strong performance on ASR tasks~\cite{hsu2021hubert, superb}.

At training time, given the paired distorted and target speech signals, we downsample the latter to 16\,kHz and extract a sequence of frozen HuBERT targets. Since HuBERT has a lower frame rate, we append an adaptive average pooling layer after the last transformer block in the speech encoder to align the sequences, and an FC layer outputs the predicted sequence. Note that although MaskSR2 restores 44.1\,kHz speech, the HuBERT targets are extracted only from 16\,kHz speech since this sampling rate covers most relevant phonetic information. Next, we consider three options to construct the HuBERT targets for semantic KD.

\vspace{1ex}
\noindent \textbf{L9-K500} quantizes the features from the 9th transformer block into 500 discrete tokens with a pre-trained k-means codebook\footref{note1}. Previous research shows that the L9 features are most effective for ASR~\cite{chen2022wavlm}, and 500 pseudo-phonetic clusters are empirically adopted by various speech tasks~\cite{zhang2023speechtokenizer, hsu2023low}. In this case, we use cross-entropy as the auxiliary loss to train the speech encoder. During inference, the HuBERT model, the pooling, and the FC layers are discarded, and there is no need to run sampling to generate the HuBERT tokens.

\vspace{1ex}
\noindent \textbf{L9-feature} refers to the continuous L9 features without quantization, which contains phonetic and presumably other helpful information for the SR task, such as speaker variability, prosody, etc. We use the mean squared error (MSE) loss in this case.

\vspace{1ex}
\noindent \textbf{Avg-feature} is the simple average of the continuous features from all the 12 transformer blocks. As shown in~\cite{chen2022wavlm}, features from different layers are good at different tasks. Though we focus on ASR, the full set of features may benefit the generated speech quality. Here we also use the MSE loss.

\vspace{1ex}
Both the continuous L9 and the average features have 768~channels. To ensure proper convergence, we perform per-channel mean and variance normalization on the target features, with the statistics extracted from each training sample. Also note that, in addition to semantic KD, the speech encoder also learns robust features from the clean HuBERT targets. To show that such robustness is not the key to the improved intelligibility, we compare semantic KD with regression to the STFT-44.1k and STFT-16k spectral targets. Both are power-law compressed magnitude STFTs $Y^{0.3}$ for a target speech signal $y$ sampled at 44.1\,kHz and 16\,kHz, respectively. Since the HuBERT targets are extracted from 16\,kHz speech, the \mbox{STFT-16k} version forms a more fair comparison. We compute 2048-point STFT for a 44.1\,kHz target signal, resulting in 1025 frequency bins, and use the first 372 bins to obtain the 16\,kHz target. The same MSE loss and feature normalization are used.

\begin{table*}[t]
    \caption{Full-band SR results on the VCTK test set. If not stated, guidance $w=1$. Bold results are the winners in each group.}
    \centering
    \resizebox{\textwidth}{!}{
    \begin{tabular}{@{}cccccccccc@{}}
    \toprule
    \multirow{2}{*}{System} & \multirow{2}{*}{Model size} & \multirow{2}{*}{KD type}
     & \multicolumn{3}{c}{DNSMOS $\uparrow$} & \multirow{2}{*}{SESQA $\uparrow$} & \multirow{2}{*}{LSD $\downarrow$} & \multirow{2}{*}{WER (\%) $\downarrow$} &
     \multirow{2}{*}{Spk Sim $\uparrow$} \\
    \cmidrule(lr){4-6}
     & & & SIG & BAK & OVL & & & & \\
    \midrule
    Unprocessed & - & - & 2.943 & 2.945 & 2.404 & 2.541 & 1.917 & 6.52 & 0.898 \\
    Target & - & - & \textbf{3.488} & 4.042 & \textbf{3.186} & \textbf{3.554} & \textbf{0.000} & 0.55 & \textbf{1.000} \\
    Target-DAC & - & - & 3.465 & \textbf{4.043} & 3.164 & 3.535 & 0.867 & \textbf{0.50} & 0.931 \\
    \midrule
    DFNet3 & 2.3\,M & - & 3.266 & \textbf{4.066} & 2.981 & 3.115 & 1.899 & \textbf{8.64} & \textbf{0.880} \\
    VoiceFixer & 111\,M & - & \textbf{3.387} & 4.014 & \textbf{3.090} & \textbf{3.330} & \textbf{1.106} & 10.16 & 0.768 \\
    \midrule
    MaskSR-S ($w=0$) & 55\,M & None & 3.419 & \textbf{4.107} & 3.159 & 3.524 & 1.257 & 8.70 & 0.818 \\
    MaskSR-S ($w=1$) & 55\,M & None & \textbf{3.459} & 4.033 & \textbf{3.160} & \textbf{3.536} & \textbf{1.163} & \textbf{6.46} & \textbf{0.842} \\
    \midrule
    MaskSR2-S & 55\,M & STFT-44.1k & 3.457 & 4.032 & 3.155 & 3.539 & \textbf{1.077} & 5.41 & 0.846 \\
    MaskSR2-S & 55\,M & STFT-16k & 3.453 & 4.010 & 3.142 & 3.536 & 1.100 & 5.59 & 0.843 \\
    MaskSR2-S & 55\,M & L9-K500 & 3.456 & 4.031 & 3.160 & 3.548 & 1.102 & 4.06 & 0.842 \\
    MaskSR2-S & 55\,M & L9-feature & 3.451 & \textbf{4.045} & 3.158 & 3.534 & 1.108 & 4.49 & 0.838 \\
    MaskSR2-S & 55\,M & Avg-feature & \textbf{3.466} & 4.038 & \textbf{3.169} & \textbf{3.561} & 1.091 & \textbf{4.01} & \textbf{0.846} \\
    \midrule
    MaskSR-L & 249\,M & None & 3.476 & 4.034 & 3.174 & 3.527 & 1.084 & 4.74 & 0.859 \\
    MaskSR2-L & 249\,M & Avg-feature & \textbf{3.490} & \textbf{4.048} & \textbf{3.195} & \textbf{3.547} & \textbf{1.065} & \textbf{3.18} & \textbf{0.861} \\
    \bottomrule
    \end{tabular}
    }
    \label{tab:fullband_results}
\end{table*}

\subsection{Audio Tokenizer and Masked Acoustic Modeling}
\label{sec:remain}

These components remain the same as in MaskSR. We use the publicly available pre-trained Descript Audio Codec (DAC)~\cite{kumar2024high} as the audio tokenizer. In the training stage, DAC tokenizes a 44.1\,kHz target speech signal to a $9\times{T}$ codegram using 9 residual vector quantizers, each with a codebook size 1024. The frame spacing of the codegram is consistent with that of the speech encoder representation of the corrupted speech. At inference time, DAC converts codegrams to waveforms. DAC is pre-trained to accurately reconstruct the unquantized waveform. Thus, the discrete DAC tokens encode acoustic details of the target speech.

The generative model leverages the MaskGIT paradigm~\cite{chang2022maskgit}. During training, a random subset of the tokens in the target codegram are masked by a special token, and 9 embedding tables embed the 9 codebook sequences, respectively, with the resulting embeddings summed to a $T\times{d}$ tensor. This tensor is also summed with the learned semantic features from the last transformer block of the speech encoder. We then process the aggregated sequence with a stack of self-attention transformer blocks. An output layer with 9 1024-channel softmax classifiers predicts the logit scores of the tokens. The speech encoder and the MaskGIT are jointly optimized by the unweighted sum of the semantic KD loss (depending on the variants in Sec.~\ref{sec:encoder}) and the cross-entropy loss on the masked positions.

During inference, starting from a fully masked codegram, the target codegram is iteratively synthesized. In each iteration, given the tokens generated from previous iterations, we sample from the output distributions in all the masked positions to unmask these tokens, and we re-mask the sampled tokens with low logit scores. The percentage of re-masking is controlled by a cosine schedule. Gaussian noise is added to the logit scores before ranking them to prevent the generated speech from being too dry~\cite{lezama2022improved}. The variance linearly decreases from 4 to 0 as a function of the iteration. In addition, we find that the classifier-free guidance~\cite{ho2022classifier} used in MaskSR is important for reducing WER (not examined previously in MaskSR). During training, the speech encoder
output is randomly replaced with a learnable embedding 10\,\% of the time, and during inference a larger guidance level $w\ge0$ conditions the generation more closely on the corrupted speech\footnote{The inference time logit scores $l_g$ are computed as $l_g = (1+w)l_c - wl_u$ where $l_c$ and $l_u$ are the conditional and unconditional logits, respectively.}, thus preserving original phonemes. We optimize $w$ on a held-out dev set for each model independently, and observe that $w=1$ generally provides the best trade-off between intelligibility and quality, whereas a larger $w>1$ does not further reduce the WER but hurts the quality. Thus, we use $w=1$ to report the results.

\section{Experimental Setup}
\label{sec:setup}

\subsection{Datasets}
\label{sec:data}

\noindent \textbf{Training set ---} 
About 800 hours of clean speech is used, including the `read speech' subset provided by the 2022 DNS Challenge~\cite{dubey2022icassp}, VCTK~\cite{veaux2017cstr}, and AISHELL-1~\cite{bu2017aishell}. We use 181 hours of noise and 60\,k room impulse responses (RIRs), also provided by~\cite{dubey2022icassp}. All data are recorded with 48\,kHz or 44.1\,kHz sampling rates, and we downsample the 48\,kHz data to 44.1\,kHz. The corrupted speech considering noise, reverb, low bandwidth, and clipping is created on the fly using the open-source data pipeline provided by VoiceFixer~\cite{liu2022voicefixer}, resulting in distorted samples with an SNR in $[-5, 20]$\,dB, a bandwidth between $[1, 22.05]$\,kHz, and clipped between $[0.1, 0.5]$. We pre-extract from the target speech the DAC codegram and the HuBERT targets before the online data creation for efficient training.

\vspace{1ex}
\noindent \textbf{Full-band test set ---} 
To create a 44.1\,kHz test set with transcripts, we simulate 1000 samples considering all 4 studied artifacts by distorting speech with a similar pipeline. We use 8 VCTK speakers, noise samples from the VoiceBank-DEMAND dataset~\cite{valentini2016investigating}, and simulated RIRs created by the WHAMR! scripts and room settings~\cite{maciejewski2020whamr}. All the test speech, noise, and RIRs are unseen during training.    

\vspace{1ex}
\noindent \textbf{Wide-band test sets ---} 
To evaluate WER on more challenging sentences with a larger vocabulary and longer duration, we simulate 800 samples using 40 speakers in the LibriSpeech test-clean dataset~\cite{panayotov2015librispeech}. We also consider the public 2020 DNS Challenge test set~\cite{reddy2020interspeech}. All the wide-band test sets are sampled at 16\,kHz, and to fairly compare with denoising-only models, we consider the test sets with mainly additive noise (without noticeable reverb). We downsample the full-band output to 16\,kHz to compare with other models.

\subsection{Implementation Details}
\label{sec:implement}

We first train small MaskSR-S and MaskSR2-S. There are 6 and 8 transformer blocks in the speech encoder and the generative model, respectively, each with dimension $d=512$ split across 16 attention heads, an MLP with a hidden dimension of $4d$, and pre-norm. Sinusoidal positional encoding is added to both the speech encoder and the generative model inputs. Then, we train large \mbox{MaskSR-L} and \mbox{MaskSR2-L} with $d=1024$ and increase the number of transformer blocks in the generative model to 12. All models are trained on 4\,sec speech segments for 800\,k steps on 8 V100 GPUs with a learning rate of 0.0001 using the Adam optimizer. We use a batch size 128 and 256 for the small and large models, respectively. During inference, we decode each non-overlapping 4\,sec window with 20 iterations.

\subsection{Baseline Models}
\label{sec:baseline}

\noindent \textbf{Full-band models ---} 
We consider 2 full-band models: VoiceFixer~\cite{liu2022voicefixer} and DeepFilterNet3 (DFNet3)~\cite{schroter2023deepfilternet}. VoiceFixer is a representative GAN-based SR model targeting at the same 4 studied distortions. DFNet3 is a strong regression model trained to remove noise and (a small amount of) reverb and clipping. 

\vspace{1ex}
\noindent \textbf{Wide-band models ---} 
We also consider various models specialized in denoising on the 16\,kHz test sets, which contain mainly additive noise. These models include regression-based DEMUCS~\cite{defossez2020real} and FRCRN~\cite{zhao2022frcrn}, diffusion models SGMSE~\cite{richter2023speech} and StoRM~\cite{lemercier2023storm}, and language model SELM~\cite{wang2024selm}.

\vspace{1ex}
\noindent \textbf{Unprocessed, Target, Target-DAC ---} These refer to the corrupted speech, target speech, and DAC-processed target speech, respectively. Target-DAC is an upper bound for MaskSR and MaskSR2.

\vspace{1ex}
Most of the models (except for VoiceFixer) are trained on datasets comparable to ours, with the majority of the data based on DNS Challenge and VCTK. Thus, we either use their reported results or run the official checkpoints to obtain the results. The released VoiceFixer was trained on a different dataset, but re-training it on our data did not obtain better results. Thus, we stick with the official checkpoint. For SGMSE and StoRM we use the results reported in~\cite{wang2024selm}, which re-trained these models on a dataset comparable to ours.

\subsection{Evaluation Metrics}
\label{sec:metric}

\vspace{1ex}
\noindent \textbf{Quality ---} Standard metrics such as PESQ and SI-SNR may not properly capture the quality of generative models due to a lack of waveform alignment~\cite{jassim2021warp}. Instead, we use the public DNSMOS~\cite{reddy2021dnsmos} and our in-house SESQA~\cite{serra2021sesqa} as reference-free perceptual quality estimators, capable of evaluating generative models in previous works targeting at removing similar distortions~\cite{li2024masksr, serra2022universal, wang2024selm}. We also adopt the open-source log-spectral distance (LSD)~\cite{liu2022voicefixer} as a common metric to measure bandwidth extension. SESQA and LSD work with full-band speech and DNSMOS works with wide-band speech. We resample the processed speech from various models if necessary.

\vspace{1ex}
\noindent \textbf{Intelligibility ---} We rely on the WER obtained by a publicly available ASR model\footnote{\url{https://huggingface.co/nvidia/stt_en_conformer_transducer_xlarge}} to measure the speech intelligibility.

\vspace{1ex}
\noindent \textbf{Speaker Similarity ---} We also measure the speaker cosine similarity (Spk Sim) between the target and the processed speech. We use the public WeSpeaker~\cite{wang2023wespeaker} to extract the speaker embeddings.

\vspace{1ex}
\noindent \textbf{Subjective Listening ---} 
13 expert listeners rate the quality of the generated speech from anonymized full-band systems on a 1--5 scale considering the 4 studied distortions (without assessing intelligibility due to the commonly used WER). We report the mean opinion scores (MOS) based on 40 samples from the VCTK test set.

\vspace{1ex}
On the full-band VCTK test set consisting of all the studied distortions, we use all the metrics to evaluate the speech quality and intelligibility. On the wide-band test sets that contain mainly additive noise, we remove SESQA and LSD since we do not target at bandwidth extension (measured by both) and declipping (measured by SESQA). On the DNS test sets we also remove WER or additionally, speaker similarity, depending on the available ground truth.

\section{Results}
\label{sec:results}
\subsection{Full-band Speech Restoration}
\label{sec:full_restore}

\begin{table}[t]
  \caption{Subjective MOS scores with 95\,\% confidence intervals.}
  \label{tab:listen_result}
  \centering
  \resizebox{1.0\columnwidth}{!}{
  \begin{tabular}{ c  c c c c c }
    \toprule
    Unprocessed & Target & DFNet3 & VoiceFixer & MaskSR-L & MaskSR2-L \\
    \midrule
    {2.01 $\pm$ 0.07} & {4.76 $\pm$ 0.04} & {2.92 $\pm$ 0.09} & {3.14 $\pm$ 0.10} & 4.53 $\pm$ 0.06 & \textbf{4.72 $\pm$ 0.05} \\
    \bottomrule
  \end{tabular}
  }
\end{table}

\begin{table}[t]
    \caption{Wide-band denoising results on the LibriSpeech test set. Bold results are the best ones in each column.}
    \centering
    \resizebox{\columnwidth}{!}{
    \begin{tabular}{ccccccc}
    \toprule
    \multirow{2}{*}{System} & \multicolumn{3}{c}{DNSMOS $\uparrow$} & \multirow{2}{*}{WER (\%) $\downarrow$} &
     \multirow{2}{*}{Spk Sim $\uparrow$} \\
    \cmidrule(lr){2-4}
     & SIG & BAK & OVL & & & \\
    \midrule
    Unprocessed & 3.053 & 2.703 & 2.396 & 5.12 & 0.959 \\
    Target & 3.590 & 3.993 & 3.274 & 1.73 & 1.000 \\
    \midrule
    DEMUCS & 3.532 & 4.069 & 3.260 & 4.48 & 0.936 \\
    FRCRN & 3.544 & \textbf{4.110} & 3.273 & \textbf{3.18} & \textbf{0.957} \\
    \midrule
    MaskSR-S & 3.580 & 4.091 & 3.319 & 5.67 & 0.885 \\
    MaskSR2-S & 3.607 & 4.070 & 3.330 & 4.60 & 0.886 \\
    MaskSR2-L & \textbf{3.632} & 4.096 & \textbf{3.366} & 4.01 & 0.901 \\
    \bottomrule
    \end{tabular}
    }
    \label{tab:LS_results}
\end{table}

Table~\ref{tab:fullband_results} shows that DFNet3, VoiceFixer, and MaskSR-S (without guidance) obtain worse WER than that of the corrupted speech despite of the improved quality, illustrating the challenge to improve both aspects. The classifier-free guidance generally improves the MaskSR-S results, especially the WER and speaker similarity. Semantic KD substantially improves WER further. We also see that KD with the HuBERT targets is superior to STFT targets, showing the key contribution of the semantic KD relative to learning robust spectral features. The Avg-feature provides the best performance (default option hereafter), showing the benefits of using the complete information from all HuBERT layers. 
The resulting \mbox{MaskSR2-S} reduces the WER by 37.9\,\% relative to MaskSR-S. Finally, scaling MaskSR-S to MaskSR-L greatly reduces WER, and \mbox{MaskSR2-L} further achieves 32.9\,\% WER reduction relative to MaskSR-L. Noticeably, the semantic KD also generally improves other aspects of the generated speech compared with MaskSR.

Table~\ref{tab:listen_result} reports the subjective listening scores based on the 40 VCTK test samples with the 4 blended distortions. MaskSR2-L outperforms MaskSR-L, nearly achieving the target quality. This shows the effectiveness of the semantic KD on the quality aspect of the generated speech. It also significantly outperforms other models.

\subsection{Wide-band Denoising}
\label{sec:wide_denoising}

\begin{table}[t]
    \caption{Wide-band denoising results on two DNS Challenge test sets. RG, DF, and LM refer to regression, diffusion, and language models, respectively.}
    \centering
    \resizebox{\columnwidth}{!}{
    \begin{tabular}{@{}ccccccccc@{}}
    \toprule
    \multirow{3}{*}{System} & \multirow{3}{*}{Type} &
      \multicolumn{4}{c}{Without Reverb} &
      \multicolumn{3}{c}{Real Recordings} \\
    \cmidrule(lr){3-6} \cmidrule(lr){7-9}
     & & \multicolumn{3}{c}{DNSMOS $\uparrow$} & \multirow{2}{*}{Spk Sim $\uparrow$} &
     \multicolumn{3}{c}{DNSMOS $\uparrow$} \\
    \cmidrule(lr){3-5} \cmidrule(lr){7-9}
     & & SIG & BAK & OVL & & SIG & BAK & OVL \\
    \midrule
    Unprocessed & - & 3.392 & 2.618 & 2.483 & 0.969 & 3.053 & 2.509 & 2.255 \\ 
    \midrule
    DEMUCS & RG & 3.575 & 4.153 & 3.345 & 0.956 & 3.263 & 4.027 & 2.988 \\
    FRCRN & RG & 3.578 & 4.133 & 3.335 & \textbf{0.970} & 3.370 & 3.977 & 3.037 \\
    \midrule
    SGMSE & DF & 3.501 & 3.710 & 3.137 & - & 3.297 & 2.894 & 2.793 \\
    StoRM & DF & 3.514 & 3.941 & 3.205 & - & 3.410 & 3.379 & 2.940 \\
    \midrule
    SELM & LM & 3.508 & 4.096 & 3.258 & - & \textbf{3.591} & 3.435 & 3.124 \\
    \midrule
    MaskSR-S & LM & 3.616 & \textbf{4.170} & 3.382 & 0.921 & 3.444 & 4.060 & 3.170 \\
    MaskSR2-S & LM & 3.629 & 4.150 & 3.391 & 0.922 & 3.495 & 4.047 & 3.204 \\
    MaskSR2-L & LM & \textbf{3.638} & 4.153 & \textbf{3.400} & 0.930 & 3.511 & \textbf{4.078} & \textbf{3.233} \\ 
    \bottomrule
    \end{tabular}
    }
    \label{tab:wideband_results}
    \vspace{-2ex}
\end{table}

On the LibriSpeech test set that contains longer sentences built upon a much larger vocabulary than that of the VCTK test set (Table~\ref{tab:LS_results}), MaskSR2-S obtains a lower WER by 18.9\,\% relative to MaskSR-S, showing the importance of the semantic KD. \mbox{MaskSR2-L} further achieves a WER lower than DEMUCS, thus yielding competitive performance compared with strong regression models trained to optimize the waveform or spectrum alignment. For speech quality, on both the LibriSpeech and the DNS test sets which contain mainly additive noise (Tables~\ref{tab:LS_results} and~\ref{tab:wideband_results}, respectively), MaskSR2 still provides superior quality. This is despite the fact that MaskSR2 is capable of removing a diverse set of distortions and thus being less specialized. 

\section{Conclusion}
\label{sec:conclusion}

In this work, we proposed MaskSR2, an improved generative framework based on MaskSR for full-band speech restoration with improved intelligibility and quality. MaskSR2 combines semantic KD and acoustic language modeling. Compared with MaskSR, the text-free semantic KD based on SSL greatly reduces the WER of the generated speech without increasing the model size and inference time. MaskSR2 also achieves competitive WER and superior quality compared to several other models. To further improve the performance, we will explore more powerful SSL models~\cite{chen2022wavlm} and a multitask speech encoder~\cite{serra2022universal} that learns from various teacher models.

\bibliographystyle{IEEEtran.bst}
\bibliography{mybib}

\begin{thebibliography}{10}
\providecommand{\url}[1]{#1}
\csname url@samestyle\endcsname
\providecommand{\newblock}{\relax}
\providecommand{\bibinfo}[2]{#2}
\providecommand{\BIBentrySTDinterwordspacing}{\spaceskip=0pt\relax}
\providecommand{\BIBentryALTinterwordstretchfactor}{4}
\providecommand{\BIBentryALTinterwordspacing}{\spaceskip=\fontdimen2\font plus
\BIBentryALTinterwordstretchfactor\fontdimen3\font minus \fontdimen4\font\relax}
\providecommand{\BIBforeignlanguage}[2]{{%
\expandafter\ifx\csname l@#1\endcsname\relax
\typeout{** WARNING: IEEEtran.bst: No hyphenation pattern has been}%
\typeout{** loaded for the language `#1'. Using the pattern for}%
\typeout{** the default language instead.}%
\else
\language=\csname l@#1\endcsname
\fi
#2}}
\providecommand{\BIBdecl}{\relax}
\BIBdecl

\bibitem{liu2022voicefixer}
H.~Liu, Q.~Kong, Q.~Tian, Y.~Zhao, D.~Wang, C.~Huang, and Y.~Wang, ``{VoiceFixer: Toward General Speech Restoration with Neural Vocoder},'' \emph{arXiv preprint arXiv:2109.13731}, 2021.

\bibitem{serra2022universal}
J.~Serr{\`a}, S.~Pascual, J.~Pons, R.~O. Araz, and D.~Scaini, ``{Universal Speech Enhancement with Score-based Diffusion},'' \emph{arXiv preprint arXiv:2206.03065}, 2022.

\bibitem{chen2023gesper}
W.~Liu, Y.~Shi, J.~Chen, W.~Rao, S.~He, A.~Li, Y.~Wang, and Z.~Wu, ``{Gesper: A Restoration-Enhancement Framework for General Speech Reconstruction},'' in \emph{INTERSPEECH}, 2023, pp. 4044--4048.

\bibitem{koizumi2023miipher}
Y.~Koizumi, H.~Zen, S.~Karita, Y.~Ding, K.~Yatabe, N.~Morioka, Y.~Zhang, W.~Han, A.~Bapna, and M.~Bacchiani, ``{Miipher: A Robust Speech Restoration Model Integrating Self-Supervised Speech and Text Representations},'' in \emph{WASPPA}, 2023, pp. 1--5.

\bibitem{zhao2022frcrn}
S.~Zhao, B.~Ma, K.~N. Watcharasupat, and W.~Gan, ``{FRCRN: Boosting Feature Representation Using Frequency Recurrence for Monaural Speech Enhancement},'' in \emph{ICASSP}, 2022, pp. 9281--9285.

\bibitem{defossez2020real}
A.~Défossez, G.~Synnaeve, and Y.~Adi, ``{Real Time Speech Enhancement in the Waveform Domain},'' in \emph{INTERSPEECH}, 2020, pp. 3291--3295.

\bibitem{li2021simultaneous}
A.~Li, W.~Liu, X.~Luo, G.~Yu, C.~Zheng, and X.~Li, ``{A Simultaneous Denoising and Dereverberation Framework with Target Decoupling},'' in \emph{INTERSPEECH}, 2021, pp. 2801--2805.

\bibitem{wang2023speechx}
X.~Wang, M.~Thakker, Z.~Chen, N.~Kanda, S.~E. Eskimez, S.~Chen, M.~Tang, S.~Liu, J.~Li, and T.~Yoshioka, ``{SpeechX: Neural Codec Language Model as a Versatile Speech Transformer},'' \emph{arXiv preprint arXiv:2308.06873}, 2023.

\bibitem{scheibler2024universal}
R.~Scheibler, Y.~Fujita, Y.~Shirahata, and T.~Komatsu, ``{Universal Score-based Speech Enhancement with High Content Preservation},'' \emph{arXiv preprint arXiv:2406.12194}, 2024.

\bibitem{e3net}
M.~Thakker, S.~E. Eskimez, T.~Yoshioka, and H.~Wang, ``{Fast Real-time Personalized Speech Enhancement: End-to-End Enhancement Network (E3Net) and Knowledge Distillation},'' in \emph{INTERSPEECH}, 2022, pp. 991--995.

\bibitem{le2024voicebox}
M.~Le, A.~Vyas, B.~Shi, B.~Karrer, L.~Sari, R.~Moritz, M.~Williamson, V.~Manohar, Y.~Adi, J.~Mahadeokar, and W.~Hsu, ``{Voicebox: Text-guided Multilingual Universal Speech Generation at Scale},'' \emph{NeurIPS}, vol.~36, 2024.

\bibitem{li2024masksr}
X.~Li, Q.~Wang, and X.~Liu, ``{MaskSR: Masked Language Model for Full-band Speech Restoration},'' in \emph{INTERSPEECH}, 2024.

\bibitem{hsu2021hubert}
W.~Hsu, B.~Bolte, Y.~H. Tsai, K.~Lakhotia, R.~Salakhutdinov, and A.~Mohamed, ``{HuBERT: Self-supervised Speech Representation Learning by Masked Prediction of Hidden Units},'' \emph{IEEE/ACM TASLP}, vol.~29, pp. 3451--3460, 2021.

\bibitem{agostinelli2023musiclm}
A.~Agostinelli, T.~I. Denk, Z.~Borsos, J.~Engel, M.~Verzetti, A.~Caillon, Q.~Huang, A.~Jansen, A.~Roberts, M.~Tagliasacchi, M.~Sharifi, N.~Zeghidour, and C.~Frank, ``{MusicLM: Generating Music from Text},'' \emph{arXiv preprint arXiv:2301.11325}, 2023.

\bibitem{kharitonov2023speak}
E.~Kharitonov, D.~Vincent, Z.~Borsos, R.~Marinier, S.~Girgin, O.~Pietquin, M.~Sharifi, M.~Tagliasacchi, and N.~Zeghidour, ``{Speak, Read and Prompt: High-fidelity Text-to-Speech with Minimal Supervision},'' \emph{Transactions of the Association for Computational Linguistics}, vol.~11, pp. 1703--1718, 2023.

\bibitem{liu2024audioldm}
H.~Liu, Y.~Yuan, X.~Liu, X.~Mei, Q.~Kong, Q.~Tian, Y.~Wang, W.~Wang, Y.~Wang, and M.~D. Plumbley, ``{AudioLDM 2: Learning Holistic Audio Generation with Self-supervised Pretraining},'' \emph{IEEE/ACM TASLP}, 2024.

\bibitem{dong2023clipsonic}
H.~Dong, X.~Liu, J.~Pons, G.~Bhattacharya, S.~Pascual, J.~Serr{\`a}, T.~Berg-Kirkpatrick, and J.~McAuley, ``{CLIPSonic: Text-to-audio Synthesis with Unlabeled Videos and Pretrained Language-vision Models},'' in \emph{WASPAA}, 2023, pp. 1--5.

\bibitem{wang2024selm}
Z.~Wang, X.~Zhu, Z.~Zhang, Y.~Lv, N.~Jiang, G.~Zhao, and L.~Xie, ``{SELM: Speech Enhancement Using Discrete Tokens and Language Models},'' in \emph{ICASSP}, 2024, pp. 11\,561--11\,565.

\bibitem{panayotov2015librispeech}
V.~Panayotov, G.~Chen, D.~Povey, and S.~Khudanpur, ``{LibriSpeech: An ASR Corpus Based on Public Domain Audio Books},'' in \emph{ICASSP}, 2015, pp. 5206--5210.

\bibitem{superb}
S.~Yang, P.~Chi, Y.~Chuang, C.~J. Lai, K.~Lakhotia, Y.~Y. Lin, A.~T. Liu, J.~Shi, X.~Chang, G.~Lin, T.~Huang, W.~Tseng, K.~Lee, D.~Liu, Z.~Huang, S.~Dong, S.~Li, S.~Watanabe, A.~Mohamed, and H.~Lee, ``{SUPERB: Speech processing Universal PERformance Benchmark},'' in \emph{INTERSPEECH}, 2021, pp. 1194--1198.

\bibitem{chen2022wavlm}
S.~Chen, C.~Wang, Z.~Chen, Y.~Wu, S.~Liu, Z.~Chen, J.~Li, N.~Kanda, T.~Yoshioka, X.~Xiao, J.~Wu, L.~Zhou, S.~Ren, Y.~Qian, Y.~Qian, J.~Wu, M.~Zeng, X.~Yu, and F.~Wei, ``{WavLM: Large-scale Self-supervised Pre-training for Full Stack Speech Processing},'' \emph{IEEE Journal of Selected Topics in Signal Processing}, vol.~16, no.~6, pp. 1505--1518, 2022.

\bibitem{zhang2023speechtokenizer}
X.~Zhang, D.~Zhang, S.~Li, Y.~Zhou, and X.~Qiu, ``{Speechtokenizer: Unified Speech Tokenizer for Speech Large Language Models},'' in \emph{ICLR}, 2024.

\bibitem{hsu2023low}
P.~Hsu, A.~Elkahky, W.~Hsu, Y.~Adi, T.~A. Nguyen, J.~Copet, E.~Dupoux, H.~Lee, and A.~Mohamed, ``{Low-Resource Self-Supervised Learning with SSL-Enhanced TTS},'' \emph{arXiv preprint arXiv:2309.17020}, 2023.

\bibitem{kumar2024high}
R.~Kumar, P.~Seetharaman, A.~Luebs, I.~Kumar, and K.~Kumar, ``{High-fidelity Audio Compression with Improved RVQGAN},'' \emph{NeurIPS}, vol.~36, 2024.

\bibitem{chang2022maskgit}
H.~Chang, H.~Zhang, L.~Jiang, C.~Liu, and W.~T. Freeman, ``{MaskGIT: Masked Generative Image Transformer},'' in \emph{CVPR}, 2022, pp. 11\,315--11\,325.

\bibitem{lezama2022improved}
J.~Lezama, H.~Chang, L.~Jiang, and I.~Essa, ``{Improved Masked Image Generation with Token-critic},'' in \emph{European Conference on Computer Vision}, 2022, pp. 70--86.

\bibitem{ho2022classifier}
J.~Ho and T.~Salimans, ``{Classifier-free Diffusion Guidance},'' \emph{arXiv preprint arXiv:2207.12598}, 2022.

\bibitem{dubey2022icassp}
H.~Dubey, V.~Gopal, R.~Cutler, A.~Aazami, S.~Matusevych, S.~Braun, S.~E. Eskimez, M.~Thakker, T.~Yoshioka, H.~Gamper, and R.~Aichner, ``{ICASSP 2022 Deep Noise Suppression Challenge},'' in \emph{ICASSP}, 2022, pp. 9271--9275.

\bibitem{veaux2017cstr}
C.~Veaux, J.~Yamagishi, and K.~MacDonald, ``{CSTR VCTK Corpus: English Multi-speaker Corpus for CSTR Voice Cloning Toolkit},'' 2017.

\bibitem{bu2017aishell}
H.~Bu, J.~Du, X.~Na, B.~Wu, and H.~Zheng, ``{AISHELL-1: An Open-source Mandarin Speech Corpus and a Speech Recognition Baseline},'' in \emph{O-COCOSDA}, 2017, pp. 1--5.

\bibitem{valentini2016investigating}
C.~Valentini-Botinhao, X.~Wang, S.~Takaki, and J.~Yamagishi, ``{Investigating RNN-based Speech Enhancement Methods for Noise-robust Text-to-Speech.}'' in \emph{SSW}, 2016, pp. 146--152.

\bibitem{maciejewski2020whamr}
M.~Maciejewski, G.~Wichern, E.~McQuinn, and J.~Le~Roux, ``{WHAMR!: Noisy and Reverberant Single-channel Speech Separation},'' in \emph{ICASSP}, 2020, pp. 696--700.

\bibitem{reddy2020interspeech}
C.~K.~A. Reddy, V.~Gopal, R.~Cutler, E.~Beyrami, R.~Cheng, H.~Dubey, S.~Matusevych, R.~Aichner, A.~Aazami, S.~Braun, P.~Rana, S.~Srinivasan, and J.~Gehrke, ``{The INTERSPEECH 2020 Deep Noise Suppression Challenge: Datasets, Subjective Testing Framework, and Challenge Results},'' in \emph{INTERSPEECH}, 2020, pp. 2492--2496.

\bibitem{schroter2023deepfilternet}
H.~Schr{\"o}ter, T.~Rosenkranz, and A.~Maier, ``{DeepFilterNet: Perceptually Motivated Real-time Speech Enhancement},'' in \emph{INTERSPEECH}, 2023.

\bibitem{richter2023speech}
J.~Richter, S.~Welker, J.~Lemercier, B.~Lay, and T.~Gerkmann, ``{Speech Enhancement and Dereverberation with Diffusion-based Generative Models},'' \emph{IEEE/ACM TASLP}, 2023.

\bibitem{lemercier2023storm}
J.~Lemercier, J.~Richter, S.~Welker, and T.~Gerkmann, ``{StoRM: A Diffusion-based Stochastic Regeneration Model for Speech Enhancement and Dereverberation},'' \emph{IEEE/ACM TASLP}, 2023.

\bibitem{jassim2021warp}
W.~A. Jassim, J.~Skoglund, M.~Chinen, and A.~Hines, ``{WARP-Q: Quality Prediction for Generative Neural Speech Codecs},'' in \emph{ICASSP}, 2021, pp. 401--405.

\bibitem{reddy2021dnsmos}
C.~K.~A. Reddy, V.~Gopal, and R.~Cutler, ``{DNSMOS: A Non-intrusive Perceptual Objective Speech Quality Metric to Evaluate Noise Suppressors},'' in \emph{ICASSP}, 2021, pp. 6493--6497.

\bibitem{serra2021sesqa}
J.~Serr{\`a}, J.~Pons, and S.~Pascual, ``{SESQA: Semi-supervised Learning for Speech Quality Assessment},'' in \emph{ICASSP}, 2021, pp. 381--385.

\bibitem{wang2023wespeaker}
H.~Wang, C.~Liang, S.~Wang, Z.~Chen, B.~Zhang, X.~Xiang, Y.~Deng, and Y.~Qian, ``{Wespeaker: A Research and Production Oriented Speaker Embedding Learning Toolkit},'' in \emph{ICASSP}, 2023, pp. 1--5.

\bibitem{kenton2019bert}
J.~Devlin, M.~Chang, K.~Lee, and K.~Toutanova, ``{BERT: Pre-training of Deep Bidirectional Transformers for Language Understanding},'' in \emph{Proceedings of naacL-HLT}, vol.~1, 2019, p.~2.

\bibitem{ziv2024masked}
A.~Ziv, I.~Gat, G.~L. Lan, T.~Remez, F.~Kreuk, A.~D{\'e}fossez, J.~Copet, G.~Synnaeve, and Y.~Adi, ``{Masked Audio Generation Using a Single Non-Autoregressive Transformer},'' \emph{arXiv preprint arXiv:2401.04577}, 2024.

\end{thebibliography}

~\newline
~\newline
~\newline

\appendix
\begin{table*}[t]
    \caption{Full-band SR results on the VCTK test set with span masking. Bold results are the winners in each group.}
    \centering
    \resizebox{\textwidth}{!}{
    \begin{tabular}{@{}cccccccccc@{}}
    \toprule
    \multirow{2}{*}{System} & \multirow{2}{*}{Span length (frames)} & \multirow{2}{*}{KD type}
     & \multicolumn{3}{c}{DNSMOS $\uparrow$} & \multirow{2}{*}{SESQA $\uparrow$} & \multirow{2}{*}{LSD $\downarrow$} & \multirow{2}{*}{WER (\%) $\downarrow$} &
     \multirow{2}{*}{Spk Sim $\uparrow$} \\
    \cmidrule(lr){4-6}
     & & & SIG & BAK & OVL & & & & \\
    \midrule
    Unprocessed & - & - & 2.943 & 2.945 & 2.404 & 2.541 & 1.917 & 6.52 & 0.898 \\
    Target & - & - & \textbf{3.488} & 4.042 & \textbf{3.186} & \textbf{3.554} & \textbf{0.000} & 0.55 & \textbf{1.000} \\
    Target-DAC & - & - & 3.465 & \textbf{4.043} & 3.164 & 3.535 & 0.867 & \textbf{0.50} & 0.931 \\
    \midrule
    MaskSR2-S & 1 & L9-K500 & \textbf{3.456} & \textbf{4.031} & \textbf{3.160} & 3.548 & 1.102 & 4.06 & \textbf{0.842} \\
    MaskSR2-S & 5 & L9-K500 & 3.453 & 4.027 & 3.157 & \textbf{3.550} & \textbf{1.101} & \textbf{3.65} & 0.839 \\
    \midrule
    MaskSR2-S & 1 & Avg-feature & 3.466 & 4.038 & 3.169 & 3.561 & 1.091 & \textbf{4.01} & \textbf{0.846} \\
    MaskSR2-S & 4 & Avg-feature & 3.460 & 4.031 & 3.162 & 3.565 & \textbf{1.086} & 4.22 & 0.844 \\
    MaskSR2-S & 6 & Avg-feature & \textbf{3.471} & \textbf{4.045} & \textbf{3.182} & \textbf{3.589} & 1.114 & 4.14 & 0.840 \\
    MaskSR2-S & 8 & Avg-feature & 3.468 & 4.039 & 3.174 & 3.570 & 1.104 & 4.11 & 0.842 \\
    \bottomrule
    \end{tabular}
    }
    \label{tab:span}
\end{table*}

\subsection{Embedding Table Initialization}
\label{sec:init}

DAC consists of 9 residual vector quantization blocks, each formed by an input FC layer that projects a 1024-dim feature vector to 8 channels, an embedding table with 1024 8-dim entries that quantizes the projected features, and an output FC layer that projects the quantized centroids back to 1024 channels. Accordingly, each embedding block in the language model also consists of an embedding table and an output FC layer with the same structure as the corresponding DAC vector quantization block (without the input projection). We find that initializing the embedding tables and the output projection weights and biases from the corresponding DAC pre-trained components improves the convergence speed to a lower loss when training MaskSR and MaskSR2. The masked token embedding (the 1025th entry) is initialized from scratch, and we take the first $d$ channels after the output projection if $d<1024$.

\subsection{Span Masking}
\label{sec:span}

To further reduce the WER, we also explored the span masking mechanism, which is widely used in masking-based SSL models~\cite{hsu2021hubert, kenton2019bert}. By masking a contiguous span of tokens (rather than single tokens) in a sequence, the model has to learn the long term dependency between the masked and unmasked tokens (typically apart by a couple of phonemes) in order to predict the masked phonetic content. Therefore, we hypothesize that span masking could be helpful to further reduce the WER of the generated speech.

Inspired by MAGNET~\cite{ziv2024masked} which shows the effectiveness of span masking in music synthesis, we extend its masking strategy to our use case. For each training sample, MAGNET randomly selects a codebook sequence $k\in\{1...9\}$ (if DAC is used), and puts spans of masks in this codebook sequence to achieve a given target mask ratio $0<r_k<1$. Meanwhile, it does not mask the lower level codebooks, and masks all the higher level codebook sequences. The model is trained to predict the masked tokens only in codebook $k$. Different than the hierarchical codebook modeling in MAGNET, we aim to achieve a global mask ratio $r_g$ by masking all codebooks with spans. Therefore, given a $9\times{T}$ codegram, we first create a $9\times{T}$ token level binary tensor that randomly masks $9\cdot{T}\cdot{r_g}$ positions in the entire codegram (but without applying this mask at this point). Then, we obtain the codebook level mask ratio $r_k$ by counting the number of masked positions in each codebook sequence. Finally, given $r_k$, $T$, and the pre-defined span length $l$, we adopt the method in MAGNET to compute the number of spans in each codebook sequence to achieve $r_k$, and mask the spans of tokens. The rest of the training pipeline is identical to that described in Sec.~\ref{sec:remain}. The inference is also based on spans. In each iteration, after obtaining the token level logit scores predicted by the language model, we compute the span level scores by taking the maximum token score in each non-overlapping span, then add Gaussian noise and rank the span level scores to re-mask the spans with low confidence.

Table~\ref{tab:span} reports the results on the full-band VCTK test set. We do not observe consistent WER reduction using span masking. For the L9-K500 setting, span masking with a span length of 5 frames brings 10.1\,\% WER reduction relative to the token level masking equivalent to using a span length of 1 frame. But for the Avg-feature case, we do not obtain WER improvements after trying various span lengths. We also observe that span masking provides comparable speech quality to that of the token level masking. We will further explore improvements to the masking strategy in our future work.

\end{document}